\documentclass[aps,prd,twocolumn,nofootinbib,superscriptaddress,preprintnumbers,a4paper,10pt]{revtex4}

\usepackage{amsmath}
\usepackage{graphicx}
\usepackage{dcolumn}
\usepackage{bm}
\usepackage{amssymb}
\usepackage{latexsym}

\begin{document}

\preprint{}

\title{Viscous Ricci Dark Energy}

\author{Chao-Jun Feng}
\email{fengcj@itp.ac.cn} \affiliation{Shanghai United Center for Astrophysics(SUCA), Shanghai Normal University, 100
Guilin Road, Shanghai 200234,China} \affiliation{Key Laboratory of Frontiers in Theoretical Physics£¬Institute of
Theoretical Physics£¬Chinese Academy of Sciences, Beijing 100190, P.R.China}
\author{Xin-Zhou Li}
\email{kychz@shnu.edu.cn} \affiliation{Shanghai United Center for Astrophysics(SUCA), Shanghai Normal University, 100
Guilin Road, Shanghai 200234,China}


\begin{abstract}
We investigate the viscous Ricci dark energy (RDE) model by assuming that there is bulk viscosity in the linear
barotropic fluid and the RDE. In the RDE model without bulk viscosity, the universe is younger than some old objects at
some redshifts. Since the age of the universe should be longer than any objects in the universe, the RDE model suffers
the age problem, especially when we consider the object APM $08279+5255$ at $z=3.91$, whose age is $t = 2.1$ Gyr. In
this letter, we find that once the viscosity is taken into account, this age problem is alleviated.

\end{abstract}


\maketitle

Recent observations like CMB anisotropy, supernovae and galaxies clustering have strongly indicated that our universe
is spatial flat and there exists a exotic cosmic fluid called dark energy with negative pressure, which constitute
about two thirds of the total energy of the universe. The dark energy is characterized by its equation of state $w$,
which lies very close to $-1$, probably being below $-1$ indicated by the present data.

Many candidates including the cosmological constant, quintessence, phantom, quintom, holographic dark energy, etc. have
been proposed to explain the acceleration. However, people still do not understand what's dark energy so far. Ricci
dark energy, which is a kind of holographic dark energy \cite{Li:2004rb} taking the square root of the inverse Ricci
scalar as its infrared cutoff, has been proposed by Gao et al.\cite{Gao:2007ep}, and this model is also
phenomenologically viable. Assuming the black hole is formed by gravitation collapsing of the perturbation in the
universe, the maximal black hole can be formed is determined by the casual connection scale $R_{CC}$ given by the
"Jeans" scale of the perturbations. For tensor perturbations, i.e. gravitational perturbations, $R_{CC}^{-2} = Max(\dot
H + 2H^2 , -\dot H )$ for a flat universe, where $H = \dot a/a$ is the Hubble parameter, and according to the
ref.\cite{Cai:2008nk}, only in the case of $R_{CC}^{-2} =\dot H + 2H^2$,  it could be consistent with the current
cosmological observations when the vacuum density appears as an independently conserved energy component. As we know,
in flat FRW universe, the Ricci scalar is $R=6(\dot H + 2H^2)$, which means the $R_{CC} \propto R$ and if one choices
the casual connection scale $R_{CC}$ as the IR cutoff, the Ricci dark energy model is also obtained. For recent
progress on Ricci dark energy and holographic dark energy, see ref.\cite{recentRicci}\cite{pert}\cite{holo:recent}. The
energy density of RDE in flat universe reads
\begin{equation}\label{energy density}
    \rho_{R} = \frac{\alpha}{2}R = 3\alpha \left(\dot H + 2H^2 \right) \,,
\end{equation}
where  we have set $8\pi G = 1$ and $\alpha$ is a dimensionless parameter which will determine the evolution behavior
of RDE.

Dissipative processes in the universe including bulk viscosity, shear viscosity and heat transport have been
conscientiously studied\cite{barrow}. The general theory of dissipation in relativistic imperfect fluid was put on a
firm foundation by Eckart\cite{eckart}, and, in a somewhat different formulation, by Landau and Lifshitz\cite{landau}.
This is only the first order deviation from equilibrium and may has a causality problem, the full causal theory was
developed by Isreal and Stewart\cite{israel}, and has also been studied in the evolution of the early
universe\cite{harko}. However, the character of the evolution equation is very complicated in the full causal theory.
Fortunately, once the phenomena are quasi-stationary, namely slowly varying on space and time scale characterized by
the mean free path and the mean collision time of the fluid particles, the conventional theory is still valid. In the
case of isotropic and homogeneous universe, the dissipative process can be modeled as a bulk viscosity $\zeta$ within a
thermodynamical approach, while the shear viscosity $\eta$ can be neglected, which is consistent with the usual
practice\cite{brevik}. Works on viscous dark energy models see ref.\cite{vde}.

The bulk viscosity introduces dissipation by only redefining the effective pressure, $p_{eff}$, according to
$p_{eff}=p-3\zeta H$ where $\zeta$ is the bulk viscosity coefficient and $H$ is the Hubble parameter. The condition
$\zeta>0$ guaranties a positive entropy production, consequently, no violation of the second law of the
thermodynamics\cite{zimdahl}. In this letter, we are interested  when the universe is dominated by a usual fluid and
the RDE, and both of them have a bulk viscosity. The case $\zeta = \sqrt{3}\tau H$, implying the bulk viscosity is
proportional to the fluid's velocity vector, is physical natural, and has been considered earlier in  a astrophysical
context, see the review article of Gr{\o}n\cite{gron}.

First, let us consider a simple case when the universe is dominated by RDE only, then the Friedmann equation reads:
\begin{equation}\label{Frid1}
    H^2 = \frac{\rho_R }{3} = \alpha \bigg[  \frac{(H^2)'}{2}  + 2H^2 \bigg]\,,
\end{equation}
where prime denotes the derivative with respect to $x \equiv \ln a$ and hereafter we set $a_0 = 1$. The solution of the
above equation is $H^2 = H_0^2e^{-2\left(2-\frac{1}{\alpha}\right)x}$ and the energy density of RDE is $\rho_R  =
3H_0^2e^{-2\left(2-\frac{1}{\alpha}\right)x}$. By using the conservation equation
\begin{equation}\label{conser1}
    \rho_R' + 3\bigg(\rho_R + p_R - 3\zeta H\bigg) = 0
\end{equation}
we obtain the pressure of viscous RDE: $P_R =
\left(1-\frac{2}{\alpha}\right)H_0^2e^{-2\left(2-\frac{1}{\alpha}\right)x} + 3\zeta H$ and the equation of state:
\begin{equation}\label{eos1}
    w = -\frac{1}{3}\left(\frac{2}{\alpha}-1\right) + \frac{\zeta}{H} \,.
\end{equation}
If we choose $\zeta = \tau \rho^s$, then
\begin{equation}\label{eos1}
    w = -\frac{1}{3}\left(\frac{2}{\alpha}-1\right) + 3^s \tau H_0^{2s-1}e^{-\left(2s-1\right)\left(2-\frac{1}{\alpha}\right)x}\,.
\end{equation}
In the special case of  $\zeta(t) = \zeta_0 = const$, i.e. $s = 0$, $w$ would be very large at very early time ($\alpha
< 1/2$) or in the later time ($\alpha >1/2$). In particular, for the case $s = 1/2$, it requires $\alpha <
2/(2+\sqrt{3}\tau)$ to accelerate the unverse ($w<-1/3$).

If the universe also contain another component with viscous RDE, the Friedmann equation and the corresponding equation
of motion can be written as:
\begin{eqnarray}
  H^2 &=& \frac{1}{3}(\rho_R + \rho_\gamma) \label{Frid2}\\
  \rho_\gamma' &=& -3(\rho_\gamma + p_\gamma - 3\zeta_\gamma H)\\
  \rho_R' &=& -3(\rho_R + p_R - 3\zeta_R H) \label{conser2}
\end{eqnarray}
where $\rho_\gamma$ is the density of fluid with a barotropic equation of state $p_\gamma = (\gamma - 1)\rho_\gamma$,
and $0\le \gamma \le 2$ to satisfy the dominant energy condition (DEC). So, the equation of state parameter $w_\gamma =
\gamma - 1$. Here we will choice $\zeta_R = \tau_R \sqrt{\rho} = \sqrt{3}\tau_RH$ and $\zeta_\gamma = \tau_\gamma
\sqrt{\rho}=\sqrt{3}\tau_\gamma H$, where $\rho = \rho_\gamma + \rho_R$. Then the Friedmann equation (\ref{Frid2})
becomes
\begin{equation}\label{Frid3}
    \alpha y'' + \left(4\alpha +3\alpha\gamma-2\right)y' + 6\left(-\gamma + 2\alpha \gamma + \sqrt{3}\tau_\gamma\right)y = 0
\end{equation}
where we have defined $y = E^2 \equiv H^2/H_0^2$. The general solution of eq.(\ref{Frid3}) is
\begin{equation}\label{gen sol}
    y = C_1 e^{-\lambda^{+}x}+ C_2 e^{-\lambda^{-}x}
\end{equation}
where  $\lambda^{\pm}=2 + \frac{3}{2}\gamma - \frac{1}{\alpha} \pm \Delta$ and $C_1$ , $C_2$ are integration constants.
Here we have defined
\begin{equation}
    \Delta = \frac{1}{2\alpha}\sqrt{\left(2-4\alpha + 3\alpha\gamma\right)^2-24\sqrt{3}\alpha \tau_\gamma} \,.
\end{equation}
From eq.(\ref{gen sol}), the background evolution only depends on $\tau_\gamma$ and does not depend on the viscosity of
RDE. By the definition of $y$, we have $C_2 = 1-C_1$, and the energy density of the fluid is
\begin{equation}
    \rho_\gamma = \frac{3}{4}H_0^2e^{-\lambda^{+}x}
  \bigg( C_1\Sigma^{+} + C_2\Sigma^{-}e^{2\Delta x}\bigg)\,,
\end{equation}
where $\Sigma^{\pm} = 2-4\alpha + 3\alpha\gamma \pm 2\alpha\Delta$. Since definition of density parameter for the fluid
is $\Omega_{\gamma0} \equiv \rho_{\gamma0}/(3H_0^2)$, then
\begin{equation}\label{c1}
    C_1 = 1-C_2 = \frac{\Omega_{\gamma0}}{\alpha \Delta} - \frac{\Sigma^{-}}{4\alpha\Delta}\,.
\end{equation}
The energy density of viscous RDE is
\begin{equation}
    \rho_R = \frac{3}{4}H_0^2e^{-\lambda^{+}x}
  \bigg[ C_1 \Pi^{-}+ C_2\Pi^{+}e^{2\Delta x}\bigg]\,,
\end{equation}
where $\Pi^{\pm} = 2+4\alpha - 3\alpha\gamma \pm 2\alpha\Delta$. By using eq.(\ref{conser2}), we obtain the equation of
state for viscous RDE:
\begin{eqnarray}
 \nonumber
  w_R \equiv \frac{p_R}{\rho_R} &=& -1 + \frac{C_1(12\sqrt{3}\tau_R + \lambda^{+}\Pi^{-})}
    {3(C_1\Pi^{-} + C_2\Pi^{+}e^{2\Delta x})}\\
   &+& \frac{C_2(12\sqrt{3}\tau_R + \lambda^{-}\Pi^{+})}
    {3(C_1\Pi^{-} + C_2\Pi^{+}e^{2\Delta x})}e^{2\Delta x}\label{eos2}\,.
\end{eqnarray}
Obviously, the present value of $w_R$ is
\begin{equation}
    w_{R0} = -1 + \frac{12\sqrt{3}\tau_R + C_1 \lambda^{+}\Pi^{-}+ C_2\lambda^{-}\Pi^{+}}{12\left(1-\Omega_{\gamma0}\right)}
\end{equation}
where we have used eq.(\ref{c1}) and if $\Delta$ is real, the past and future value of $w_R$ is
\begin{eqnarray}
  w_R(x\rightarrow -\infty) &=& -1 + \frac{\lambda^{+}}{3} + \frac{4\sqrt{3}\tau_R}{\Pi^{-}}\\
  w_R(x\rightarrow \infty) &=& -1 + \frac{\lambda^{-}}{3} + \frac{4\sqrt{3}\tau_R}{\Pi^{+}}\,,
\end{eqnarray}
and one can see that the value of the equation of state parameter is determined by both the viscosity of the fluid and
that of RDE and does not blow up neither in the past nor in the future.

In the following, we will consider the case of $\gamma = 1$, which corresponding the equation of state of dark matter
$w_m = 0$. By assuming $\tau_m \ll (2-\alpha)^2/(24\sqrt{3}\alpha) $, we obtain
\begin{eqnarray}
  \Delta &\approx& \frac{2-\alpha}{2\alpha}-\tilde\tau_m \,,\\
  \lambda^{+} &\approx& 3-\tilde\tau_m  \,,\\
  \lambda^{-} &\approx& 4-\frac{2}{\alpha}+\tilde\tau_m \,,\\
  \Sigma^{+} &\approx& 4-2\alpha - 2\alpha\tilde\tau_m \,,\\
  \Sigma^{-} &\approx& 2\alpha\tilde\tau_m \,,\\
  \Pi^{+} &\approx& 4-2\alpha\tilde\tau_m \,,\\
  \Pi^{-} &\approx& 2\alpha + 2\alpha\tilde\tau_m \,,\\
  C_1 = 1-C_2 &\approx& \frac{2\Omega_{m0}}{2-\alpha} + \frac{\alpha\tilde\tau_m}{2-\alpha}\left( \frac{4\Omega_{m0}}{2-\alpha} - 1 \right)\,,
\end{eqnarray}
where $\tilde\tau_m = \frac{6\sqrt{3}\tau_m}{2-\alpha}$. And the present equation of state is
\begin{equation}
    w_{R0} \approx -\frac{1}{3}\left[\frac{2}{\alpha}-\frac{1+3\sqrt{3}(\tau_R+\tau_m)}{1-\Omega_{m0}}\right]\,,
\end{equation}
so it requires
\begin{equation}
\alpha < \frac{2(1-\Omega_{m0})}{2-\Omega_{m0}+3\sqrt{3}(\tau_R+\tau_m)}
\end{equation}
to accelerate the universe at precent. The past and future values of the equation of state parameter are:
\begin{eqnarray}
  w_R(x\rightarrow -\infty) &\approx& \frac{2\sqrt{3}}{\alpha} \left[\tau_R - \frac{\alpha }{2-\alpha}\tau_m\right]\,,\\
  w_R(x\rightarrow \infty) &\approx& -\frac{1}{3}\left(\frac{2}{\alpha}-1\right) \\
  &+& 2\sqrt{3}\left(\frac{\tau_R}{2}+ \frac{\tau_m}{2-\alpha}\right)\,,
\end{eqnarray}
where we have assumed $\tau_R$ is the same order as $\tau_\gamma$ and kept only linear terms of them.

The age of our universe at redshift $z$ is given by $t(z) = T(z)/H_0$, where
\begin{equation}\label{age}
    T(z)=\int^\infty_z \frac{dz'}{(1+z')E(z)}=\int^{-\ln (1+z)}_{-\infty} \frac{dx}{E(x)}\,,
\end{equation}
is the so-called dimensionless age parameter. For a flat CDM universe dominated by matter ($\Omega_{m0} = 1$), $t_0 =
2/(3H_0)$, and according to the observations of the Hubble Space Telescope Key project, the present Hubble parameter is
constrained to be $H_0 = 9.776 h^{-1}$, $ 0.64< h <0.80$, which is consistent with the conclusions arising from
observations of the CMB and large scale structure. This gives $t_0 \approx 8\sim10$Gyr, which does not satisfy the
stellar age bound: $t_0 > 11\sim12$Gyr, namely, the age of the universe should be longer than any of others in the
universe. For $\Lambda$CMD model, in which $\Omega_{m0} \approx 0.27$, the age parameter is
\begin{equation}\label{age LCDM}
    T(z) = \int^\infty_z \frac{dz'}{(1+z')\left[\Omega_{m0}(1+z)^3 + (1-\Omega_{m0})\right]^{1/2} }\,,
\end{equation}
hence, it easily satisfied the constraint $t_0 > 11\sim12$Gyr, because the value of $\Omega_{m0}$ is smaller than that
in CDM model, consequently, the integrand in (\ref{age LCDM}) is bigger when $z$ is large, and thus, the total age of
the universe is going to be longer in $\Lambda$CDM model. However, the age of the universe at $z = 3.91$ in
$\Lambda$CDM model is about $t(3.91) \approx 0.12H_0^{-1}\approx 1.44\sim 1.79$Gyr which is still younger than the old
object APM $08279+5255$ observed recently\cite{apm} at the same redshift with age $2.1$ Gyr. In the RDE model without
viscosity, $t(3.91) \approx 0.10H_0^{-1}\approx 1.26\sim 1.57$Gyr with $\alpha = 0.46$, so the universe is also younger
than APM $08279+5255$ .

\begin{figure}[h]
\includegraphics[width=0.4\textwidth]{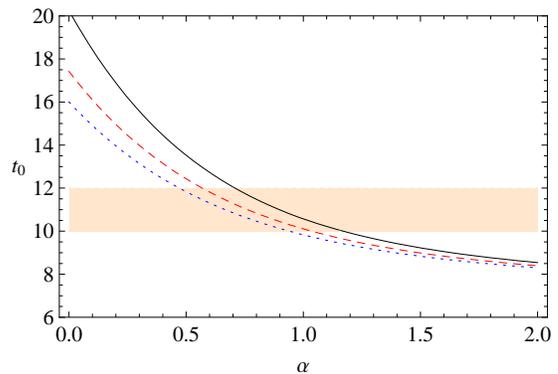}
\caption{\label{fig::Erage} The total age of the universe in RDE model with respect to parameter $\alpha$. Here $h =
0.72$ and $\Omega_{m0} = 0.20$(solid), $0.27$(dashed), $0.32$(dotted) are used. The horizonal shadow corresponds to the
stellar age bound, namely $t_0 = 10\sim12$Gyr. }
\end{figure}

\begin{figure}[h]
\includegraphics[width=0.4\textwidth]{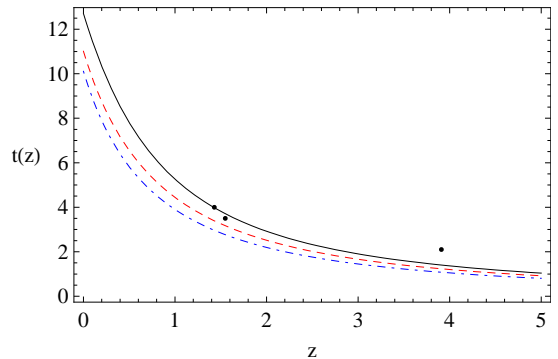}
\caption{\label{fig::Evage}The evolution of age of the universe  in RDE model with $\alpha = 0.46$(solid),
$0.76$(dashed), $1.0$(dotdashed). Here $h = 0.72$ and $\Omega_{m0} = 0.27$ are used. $\alpha < 1/3$ will make the
perturbation unstable, see \cite{pert}. }
\end{figure}

We plot the total age and its evolution with different values of $\alpha$ in Fig.\ref{fig::Erage} and
Fig.\ref{fig::Evage},
where three black points denote the age of some old objects in the universe: LBDS 53W091\cite{091} at $z=1.43$ with age
$t = 3.5$ Gyr, LBDS 53W069\cite{069} at $z=1.55$ with age $t = 4.0$ Gyr and APM $08279+5255$ at $z=3.91$ with age $t =
2.1$ Gyr. It seems that matter was diluted so fast that makes the universe younger than these old objects in the RDE
model.

\begin{figure}[h]
\includegraphics[width=0.4\textwidth]{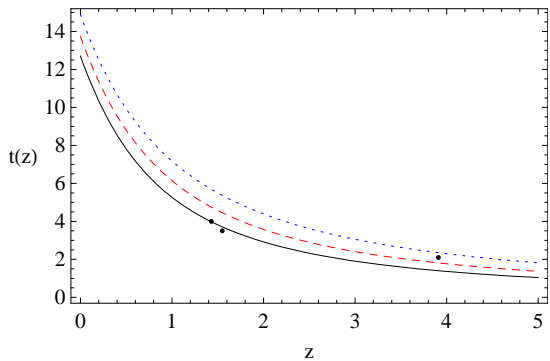}
\caption{\label{fig::vage}The evolution of age of the universe in the viscous RDE model with $\alpha = 0.46$ and
$\tau_m = 0$(solid), $0.03$(dashed), $0.06$(dotdashed). Here $h = 0.72$ and $\Omega_{m0} = 0.27$ are used. }
\end{figure}

However, this age problem can be alleviated in the viscous RDE model. We plot the  evolution of age of the universe
with different values of $\alpha$ in this model, see Fig.\ref{fig::vage} and it indicates the viscosity could really
alleviate the age problem.
And actually, only the viscosity of matter affects the evolution of age of the universe, so it could alleviate the age
problem in other cosmological models.

 In conclusion, we have investigated the Ricci dark energy model when the bulk viscosity $\zeta_R = \tau_R
\sqrt{\rho}$ is taken into account in this letter. The energy conservation equations will have additional terms
proportional to the bulk viscosity in this case. However, in this model, the evolution of the universe only depends on
the bulk viscosity $\zeta_\gamma = \tau_\gamma \sqrt{\rho}$ of ordinary fluids with equation of state $p =
(\gamma-1)\rho$, and it does not depend on $\zeta_R$. The RDE model suffers the age problem since the age of the
universe should be longer than any other objects at any redshifts in the universe. It seems that the problem is caused
by the fact that matter is diluted too fast. When one consider the  viscosity of matter, it changes the energy
conservation equation for the matter, consequently, it makes matter diluted a little bit slower, and so the age problem
is alleviated.

Considering the viscosity of fluid is a next step from the idea one to treat fluid more realized, since the real fluid
should have the viscous properties when it flows, so it is very interesting and worth further studying.

\begin{acknowledgments}
This work is supported by National Science Foundation of China grant No. 10847153 and No. 10671128. CJF would like to
thank Chang-Jun Gao, Qing-Guo Huang, Miao Li, Xin-He Meng and Shuang Wang for useful discussions.
\end{acknowledgments}

\end{document}